\documentclass[aps,twocolumn,showpacs,amsmath,amssymb,floatfix,superscriptaddress]{revtex4-1}
\usepackage{epsfig}
\usepackage{bm}
\usepackage{color}
\usepackage[hidelinks]{hyperref}

\begin{document}

\title{
Current correlations of Cooper-pair tunneling into a quantum Hall system
}

\author{Andreas B. Michelsen}
\affiliation{Department of Physics and Materials Science, University of Luxembourg, L-1511 Luxembourg, Luxembourg}
\affiliation{SUPA, School of Physics and Astronomy, University of St Andrews, North Haugh, St Andrews KY16 9SS, UK}

\author{Thomas L. Schmidt}
\affiliation{Department of Physics and Materials Science, University of Luxembourg, L-1511 Luxembourg, Luxembourg}

\author{Edvin G. Idrisov}
\affiliation{Department of Physics and Materials Science, University of Luxembourg, L-1511 Luxembourg, Luxembourg}

\date{\today}

\begin{abstract}
We study Cooper pair transport through a quantum point contact between a superconductor and a quantum Hall edge state at integer and fractional filling factors. We calculate the tunnelling current and its finite-frequency noise to the leading order in the tunneling amplitude for dc and ac bias voltage in the limit of low temperatures. At zero temperature and in case of tunnelling into a single edge channel both the conductance and differential shot noise vanish as a result of Pauli exclusion principle. In contrast, in the presence of two edge channels, this Pauli blockade is softened and a non-zero conductance and shot noise are revealed.
\end{abstract}

\maketitle

\section{Introduction}
\label{Sec:I}

The quantum Hall (QH) effect~\cite{Klitzing, Tsui} is one of the most important effects of modern mesoscopic physics. Its main observable feature is the precise quantization of the Hall conductance to the value $G_H=\nu e^2/h$, where $\nu$ is the so-called filling factor. In a two-dimensional electron gas (2DEG) at integer ($\nu \in \mathbb{N}$) or certain fractional [$\nu=1/(2n+1)$ where $n \in \mathbb{N}$] filling factors, electron transport occurs through one-dimensional (1D) channels located close at the edges of the sample~\cite{Ezawa}. The electron motion in these 1D channels is chiral, i.e., the electrons propagate in one direction with a speed of the order of $10^4$ to $10^6 \text{ m/s}$~\cite{Ashoori, Kumada, Tewari}. Electrons in such edge channels propagate ballistically without backscattering, in a way similar to photons in wave guides. This analogy has led to the emergence of the field of electron quantum optics which aims to realize quantum-optics-type experiments with electrons~\cite{Glattli1}.

Recent progress in experimental techniques at the nanoscale has allowed experimentalists to create hybrid mesoscopic systems where QH edge states are coupled to other edge states~\cite{Glattli1, Roussel, Glattli2}, to quantum dots~\cite{Tewari, Kataoka, Waintal, Meir, Fletcher}, to Ohmic contacts~\cite{Iftikhar1, Iftikhar2, Mitali1, Mitali2}, or to superconductors \cite{Moore, Rickhaus, Amet, Park, Lee}. This development has provided a successful platform to study some of the fundamental questions of mesoscopic physics, such as phase-coherence~\cite{Ji, Camino, Neder, Roche, Heiblum, Duprez1, Duprez2}, charge~\cite{Iftikhar1, Pierre1, Pierre2, Pierre3} and heat quantization~\cite{Mitali1, Pierre4}, equilibration~\cite{Pierre5, Pierre6, Pierre7, Grivnin, Fujisawa,SukhorukovNew} and entanglement~\cite{Weisz, Glattli2}. A particularly important setup for studying the transport properties of hybrid mesoscopic systems is based on QH edge states coupled to a metal via a quantum point contact (QPC), a narrow region between two electrically conducting systems. Such QPCs allow for tunneling experiments in the presence of an applied dc or ac bias voltage. In particular, the current and shot noise through a QPC connecting a QH edge state have been investigated in many experiments~\citep{Glattli1}. These experiments have made it possible to study the crossover from Fermi liquid to non-Fermi liquid phases in the $I-V$ (current-voltage) characteristics and in the corresponding noise measurements.

To study the transport in mesoscopic devices based on QH edge states, the low-energy effective theory developed by Wen is commonly used~\cite{Wen1}. This bosonization approach shows that fractional edge states of the Laughlin series [$\nu = 1/(2n + 1)$] can be modelled as Luttinger liquids with Luttinger parameter $K = \nu$. This theory has allowed the interpretation of the experimental data~\cite{Glattli1, Glattli2} obtained for transport properties of 1D chiral edge states. Moreover, the tunnelling current and conductance, as well as the zero-frequency and finite-frequency non-equilibrium noise between edge states were studied theoretically~\cite{Wen2, Chamon1, Chamon2, Chamon3, Kane1, Kane2, Martin1, Martin2, Martin3, Safi1, Safi2, Safi3, InesSafi, Ferraro, Dario2, Dario3, DarioNew, SchmidtSassetti, SchmidtDolcetto}.

In these works, it was already shown that the typical behavior of the tunneling conductance of Laughlin fractional QH chiral edge states at low temperatures follows a power law, i.e., $G(T) \propto T^{2g-2}$, where $T$ is the temperature and the parameter $g$ is equal to $\nu$ or $1/\nu$ depending on the geometry of QPC. Additionally it was shown that the behavior of the dc $I-V$ characteristic at zero temperature, low bias and $g \neq 1$ is non-Ohmic, $I_{dc}(V) \propto V^{2g-1}$, which is associated with the non-Fermi (Luttinger) liquid phase. In the case of a time-dependent bias voltage, $\tilde{V}(t)=V_0+V_1\cos(\Omega t)$ with frequency $\Omega$ and amplitude $V_1$ in the periodic ac part, the dc component of the current was found to have the form $I_{dc}=\sum_{n} J^2_n(e^{\ast} V_1/\hbar \Omega)|e^{\ast}V_0+n \hbar \Omega|^{2g-1}$, where $e^{\ast}$ is the effective charge of the tunneling particle, $J_n(e^{\ast} V_1/\hbar \Omega)$ gives the Bessel function of the first kind and $n$ is an integer number. Apart from the $I-V$ characteristic, the study of the zero- and finite-frequency noise in these references revealed a power-law dependence of the noise on the frequency at low temperatures. For instance, to the lowest order in the tunnel coupling, the finite-frequency symmetric noise at frequency $\omega$ is proportional to the sum of two terms $|\omega \pm \omega_0|^{2g-1}$, which exhibit singularities at frequencies $\omega_0=e^{\ast} V/\hbar$ and $g<1/2$. In the case of a time-dependent bias voltage, the result gets modified similarly to the current to $|\omega \pm (\omega_0+n \Omega)|^{2g-1}$, and again exhibits singularities at certain frequencies. The noise thus provides one of the most straightforward methods to measure the effective charge $e^{\ast}$ of tunneling Laughlin quasiparticles~\citep{Glattli1}.

In the recent past, it has become possible to investigate such transport problems not only between identical ballistic chiral QH states but also between distinct systems, such as QH edge states and superconductors, both theoretically~\cite{Fisher,Maslov, Fradkin,Virtanen,Hou,Hoppe,Gamayun,Cohnitz,Kim,Giazotto,Ostaay} and experimentally~\cite{Rickhaus,Amet,Lee,Chtchelkatchev,Komatsu,Stefan1,Stefan2}. This line of research is particularly relevant for the creation of parafermion bound states, non-Abelian quasiparticles with potential application in topological quantum computation \cite{lindner2012,clarke2013,klinovaja2014,alicea2016,pedder2017,calzona2018,wu2018a,groenendijk2019,schmidt2020,schiller2020}. Motivated by this progress, we investigate the noise properties of the tunneling current between a superconductor and QH edge states at integer and Laughlin filling factors. We show that the previously demonstrated Pauli blockade~\cite{Fisher} in the tunneling current at filling factor $\nu=1$ also manifests itself in shot noise experiments. We expect that one can investigate shot noise and finite-frequency noise experimentally, as was done in Refs.~[\onlinecite{Chang}] and [\onlinecite{Glattli1}] where the authors measured the dependence of noise on temperature and applied bias.

We note that due to the magnetic field the QH edge state is spinless (spin-polarised), which suppresses any induced correlations from an $s$-wave SC. This suppression can be lifted by spin-orbit coupling, such as the Rashba spin-orbit coupling inherent to the geometry of a 2DEG \cite{Ostaay}. This is relevant when the QH material is e.g. InAs, but for the popular choice of graphene this is relatively weak. It has been suggested \cite{leeInducing2017} that the QH edge state can be considered having effective spin-orbit coupling inherited through proximity with a superconductor with bulk \cite{wakamuraSpin2014} or surface \cite{kimImpurityinduced2015} spin orbit coupling.

The rest of this article is structured as follows. In Sec.~\ref{Sec:II}, we introduce the model of a QPC in the spirit of Ref.~[\onlinecite{Fisher}]. In Sec.~\ref{Sec:III}, we calculate the tunneling current and the conductance perturbatively for a finite dc bias, which we will need in the following section. In Sec.~\ref{Sec:IV}, we calculate the finite-frequency noise in the dc regime. Sec.~\ref{Sec:V} is devoted to the derivation of the tunneling current and the finite-frequency noise for a periodic ac bias voltage. Finally, we present our conclusions and some future perspectives in Sec.~\ref{Sec:VI}. Details of the calculations and additional information are presented in the Appendices. Throughout the paper, we set $|e|=\hbar=k_B=1$.

\section{Theoretical model of a quantum point contact}
\label{Sec:II}

\begin{figure}
\includegraphics[width=\columnwidth]{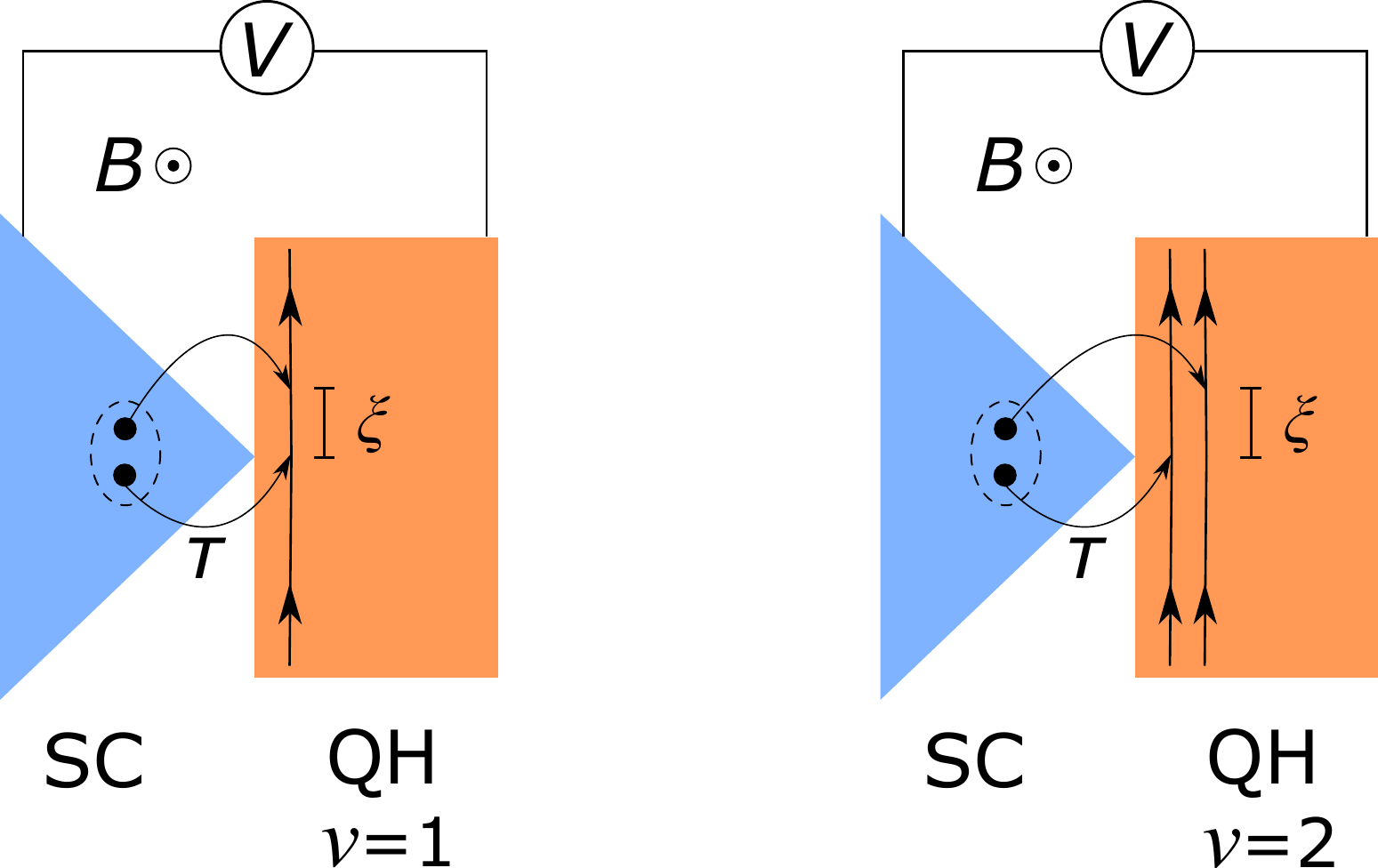}
\caption{\label{fig:QPC} Schematic representation of the system: a QPC with tunneling amplitude $\tau$ connects a superconductor (SC) to the chiral edge states of an integer quantum Hall (QH) phase at filling factor $\nu$. At $\nu=1$ both electrons of the Cooper pair would have to occupy the same state, leading to a Pauli blockade, while at $\nu=2$ the electrons can enter different states. The bias is applied between the chiral edge channel and the superconductor.}
\end{figure}

We start by introducing the Hamiltonian of a QPC between a QH edge state at filling factor $\nu$ and an $s$-wave superconductor (see Fig.~\ref{fig:QPC}). To describe this system theoretically, we use the phenomenological model presented in Ref.~\cite{Fisher}. We consider a total Hamiltonian of the system consisting of a term describing the QH edge, a term describing the SC and a tunneling term,
\begin{align}
\label{Total Hamiltonian}
\hat{H}=\hat{H}_{QH} + \hat{H}_{SC} + \hat{H}_T.
\end{align}
The exact form of the term $\hat{H}_{QH}$ depends on the filling factor and the cases of integer and fractional states as well as of a pair of co-propagating states will be presented in next sections.

Tunneling between the superconductor and the edge state is a two-step process. A Cooper pair from the superconducting condensate first splits into two electrons with opposite spins in a singlet state, both of which tunnel into the QH system. However, since the edge state is spin-polarized, a further spin-flip process, which can be brought about by spin-orbit coupling, is necessary to reach the final state which contains two electrons with the same spin propagating in the edge state. At temperatures much smaller than the superconducting gap, the Cooper pairs can be described as the mean value of the bosonic field $\hat c$ describing the superconducting condensate, $\Delta = \langle \hat c \rangle$. We assume that the Cooper pair tunneling happens at the point $x=0$, and use this to build the tunneling Hamiltonian
\begin{multline}
\hat H_T = \int dx dx^\prime\ t_1(x,x^\prime) \Big(\hat \psi^\dagger_\uparrow(x) \hat \psi^\dagger_\downarrow(x^\prime) \hat c(x=0) +\text{H.c.}\Big)
\\
+ \int dx\ t_2(x) \Big(\hat \psi^\dagger_\uparrow(x) \hat \psi_\downarrow(x) +\text{H.c.}\Big).
\end{multline}
where $t_1$ and $t_2$ are tunneling and spin-flip amplitudes, respectively, and $\psi_\uparrow(x)$ is the annihilation operator for a spin-up electron in the edge state at position $x$. If we consider the Hamiltonian perturbatively in $t_1$ and $t_2$, at second order we find the term
\begin{align}
\int dx dx^\prime\ \tau(x,x^\prime) \Big[\hat \psi^\dagger_\uparrow(x) \hat \psi^\dagger_\uparrow(x^\prime)\hat c(x=0) +\text{H.c.}\Big].
\end{align}
where we have the effective tunneling parameter $\tau(x,x^\prime) = t_1(x,x^\prime)[t_2(x) - t_2(x^\prime)]$. This term is the lowest order term which includes both spin flip and Cooper pair tunneling in such a way as to remove the Cooper pair from the SC and create two spin up electrons in the QH edge. Thus, at low energies this term will dominate the transport process across the interface, and we will neglect all other terms. The term represents an effective p-wave pairing which is suppressed at short distances by the Pauli principle and vanishes exponentially at distances larger than the superconducting coherence length $\xi \propto v_F/\Delta$ ~\cite{Varlamov}, where $v_F$ is the Fermi velocity of the SC. This allows us to effectively approximate the term using a fixed distance $\xi$ between the electrons in the final state,
\begin{align}
\label{Tunneling Hamiltonian}
\hat{H}_T^\prime = \tau \hat{\psi}_\uparrow^{\dagger}(x=\xi)\hat{\psi}_\uparrow^{\dagger}(x=0) \hat{c}(x=0)+\text{H.c.},
\end{align}
without loss of qualitative generality~\cite{Fisher}. From here on we will suppress the spin index.

In the following, we consider the effective Hamiltonian given by taking Hamiltonian~(\ref{Total Hamiltonian}) and replacing $\hat{H}_T$ with $\hat{H}_T^\prime$, which gives a complete description of the system under consideration. In the following, the relevant energy scales are assumed to be small compared to the Fermi energy, allowing us to use the effective low-energy theory to take into account the strong electron-electron interaction in edge states for the cases of filling factor $\nu=2$ and $\nu=1/(2n+1)$ ($n \in \mathbb{N}$)~\cite{Wen1,Wen2}. The tunneling term~(\ref{Tunneling Hamiltonian}) is considered perturbatively.

\section{Tunneling current in the DC regime}
\label{Sec:III}
The operator for the tunneling current is given by $\hat{J}=d\hat{N}_{QH}/dt=i[\hat{H},\hat{N}_{QH}]$, where $\hat{N}_{QH}=\int dx \hat{\psi}^{\dagger}(x)\hat{\psi}(x)$ is the electron number operator in the QH channel. It can be expressed as
\begin{equation}
\label{Tunneling current operator}
\hat{J}=2i \tau \Delta (\hat{A}^{\dagger}-\hat{A}),
\end{equation}
where the operator $\hat{A}=\hat{\psi}(0)\hat{\psi}(\xi)$ consists of two fermionic fields. According to the real-time Keldysh approach the average tunneling current in the interaction picture is given by the expression
\begin{equation}
\label{Keldysh approach}
I(t)=\langle \hat{U}^{\dagger}(t,-\infty)\hat{J}(t)\hat{U}(t,-\infty)\rangle,
\end{equation}
where the average is taken with respect to the dc biased ground state of QH edges and superconductor. The current becomes time-independent once the system has reached a steady state. At the lowest order of tunneling coupling, the time evolution operator is given by
\begin{equation}
\hat{U}(t_1,t_2) \approx 1-i\int^{t_1}_{t_2}dt \hat{H}_T(t).
\end{equation}
One then finds that the average tunneling current can be written in term of a commutator of $A$ operators~\cite{Edvin1, Edvin3, Edvin2}
\begin{equation}
\label{Tunneling current}
I_{dc}(V)=2(\tau \Delta)^2 \int_{-\infty}^\infty dt e^{2iVt} \left \langle \left[\hat{A}^{\dagger}(t), \hat{A}(0)\right] \right \rangle_0,
\end{equation}
where $V$ is the applied dc bias voltage. The average is taken with respect to the ground state of the uncoupled system, i.e., with respect to the equilibrium density matrix $\hat{\rho}_0 \propto \exp[-(\hat{H}_{QH}+\hat{H}_{SC})/T]$, where $T$ is the temperature. The integrand of Eq.~(\ref{Tunneling current}) only depends on one time variable due to time translation invariance in presence of dc bias. The pre-factor $2$ reflects the charge $2e$ of the Cooper pairs. The perturbative result is valid as long as the tunneling current is small compared to the Hall current. Restoring the natural units, the Hall current is given by the relation $I_H=\nu e^2 V/2\pi \hbar$.

\subsection{Filling factor $\nu=1$}
\label{Filling factor one}
As an illustration of our approach based on Eq.~(\ref{Tunneling current}), we first start by considering a system at filling factor $\nu=1$ and described by
\begin{equation}
\hat{H}_{QH}=-iv_F \int dx \hat{\psi}^{\dagger}(x) \partial_x \hat{\psi}(x).
\end{equation}
Without loss of generality we consider right-moving fermions and focus on a positive applied dc bias voltage $V>0$. In the case of finite temperature $T$, an analytical continuation in the complex plane is applied to Eq.~(\ref{Tunneling current}). One finds the following result for the tunneling current,
\begin{align}
\label{Tunneling current. Finite temperature. Filling factor one. DC regime}
I_{dc}(V)/I_0 &= \frac{\xi T}{v_F}\sinh\left(\frac{V}{T}\right)\\
& \times \left[\mathcal{F}\left(0,\frac{V}{\pi T}\right)-\mathcal{F}\left(\frac{2\pi \xi T}{v_F}, \frac{V}{\pi T}\right)\right] \notag,
\end{align}
where $I_0=(\tau \Delta)^2 /\pi v_F \xi$ is a normalization factor and the terms in square brackets are given by the integral
\begin{equation}
\label{Dimensionless integral F(a,b)}
\mathcal{F}(a,b)=\int_{-\infty}^{+\infty} \frac{\cos(b z)dz}{\cosh(a)+\cosh(z)}.
\end{equation}
Here one can check that the tunneling current vanishes at $V \rightarrow 0$ or $\xi \rightarrow 0$. In the general case, the result of Eq.~(\ref{Tunneling current. Finite temperature. Filling factor one. DC regime}) can be expressed in terms of hypergeometric functions. However, we are mainly interested in the regime of low temperature compared to the superconducting gap, $\xi T/v_F \ll 1$. Moreover, as we are mainly interested in the linear conductance, we also assume low voltages compared to the temperature scale, $V/ T \ll 1$. In this case the result simplifies to $I_{dc}(V) \simeq (4/3\pi)(\tau \Delta/v_F)^2 V(\pi \xi T/v_F)^2$.

\begin{figure}
\includegraphics[width=\columnwidth]{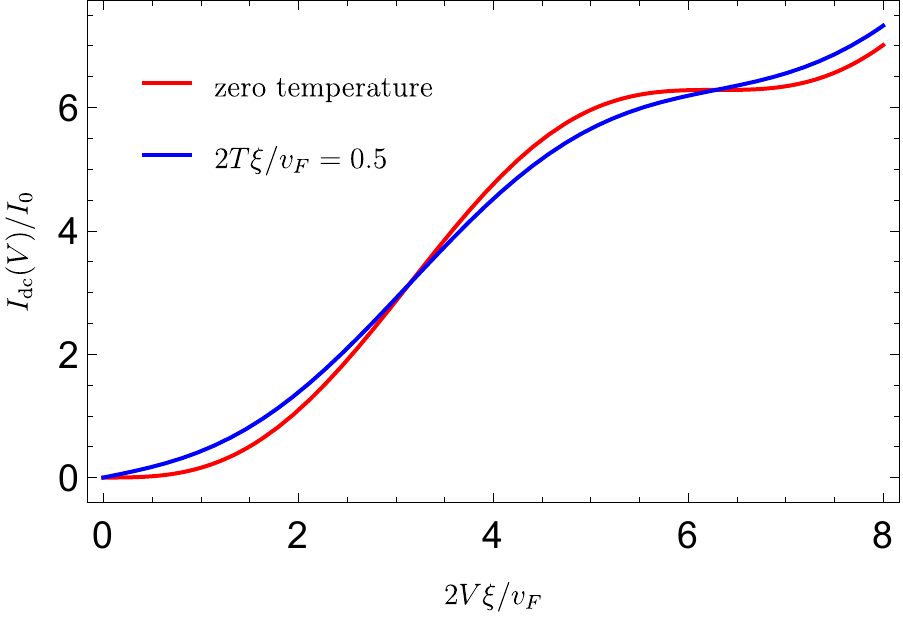}
\caption{\label{fig:two} The normalized tunneling current $I_{dc}(V)/I_0$ for filling factor $\nu=1$ oscillates with the dimensionless bias $2V\xi/v_F$ around an Ohmic behaviour, with non-zero temperature damping the oscillation (see Eqs.~(\ref{Tunneling current. Finite temperature. Filling factor one. DC regime}) and (\ref{Current. Zero temperature. Filling factor one. DC regime})).}
\end{figure}

A direct calculation of the conductance $G=\partial I_{dc} (V)/\partial V$ at $V \to 0$ from Eq.~(\ref{Tunneling current. Finite temperature. Filling factor one. DC regime}) gives
\begin{equation}
\label{Conductance. Zero temperature. Filling factor one}
G(T)/G_0=1-\frac{2\pi \xi T/v_F}{\sinh (2\pi \xi T/v_F)},
\end{equation}
where the normalization is equal to $G_0=2(\tau \Delta)^2/\pi v^2_F$. In the low-temperature limit $\xi T/v_F \ll 1$ we find that $G(T)/G_0 \simeq (2/3)(\pi \xi T/v_F)^2$.

Next, we will discuss the results at zero temperature. Using Eq.~(\ref{Tunneling current. Finite temperature. Filling factor one. DC regime}), we obtain the expression for the tunneling current at $T=0$,
\begin{equation}
\label{Current. Zero temperature. Filling factor one. DC regime}
I_{dc}(V)/I_0=  \frac{2V \xi}{v_F}\left[1-\frac{\sin(2V \xi /v_F)}{2V \xi /v_F}\right].
\end{equation}
In the limit of small bias voltage $V\xi/v_F \ll 1$, we find non-Ohmic behaviour $I_{dc}(V) \propto (\tau \Delta)^2 V^3 \xi^2/v^4_F$, as shown in Fig.~\ref{fig:two}. The oscillatory term is associated with the fact that the tunneling occurs at two points, separated by the superconducting coherence length, $\xi$. The linear QPC conductance associated with tunneling current is given by $G=\partial I_{dc} (V)/\partial V$ at $V \to 0$. The direct calculation gives $G=0$ at zero temperature. According to Ref.~[\onlinecite{Fisher}], the vanishing conductance and the non-Ohmic behavior of the tunneling current is related to the Pauli exclusion principle. At low energy scales, Pauli exclusion diminishes the effective density of states for electron-pair tunneling, namely $\rho_{DOS} \propto (V \xi/v_F)^2$ at zero temperature and $\rho_{DOS} \propto (T \xi/v_F)^2$ at finite temperature. Physically this means that after the first electron has tunneled, the tunneling of a second electron is strongly suppressed up to times $t \sim \xi/v_F$.

\subsection{Filling factor $\nu=2$}
\label{Filling factor two. DC current}
In this subsection, we consider the QH edge at filling factor $\nu=2$. First, we describe the non-interacting case. A pair of electrons from the superconductor can now tunnel simultaneously into two different edge channels~\cite{Fisher}, denoted by $1$ and $2$. To model this process, the electron operator in Eq.~(\ref{Tunneling current operator}) can be represented as a superposition of independent fermionic fields $\hat{\psi}_{1,2}(x)$ as $\hat{\psi}=\sqrt{p} \hat{\psi}_1+\sqrt{1-p}\hat{\psi}_2$, where $p$ is the probability of an electron tunneling into edge state $1$, and $1-p$ is the probability of tunneling into edge state $2$. To calculate the tunneling current~(\ref{Tunneling current}) we need the two-point correlation functions $G_j(x_1-x_2;t_1-t_2)=\langle \hat{\psi}^{\dagger}_j(x_1,t_1)\hat{\psi}_j(x_2,t_2) \rangle_0 $, where $j=1,2$ denotes the edge channel and for simplicity we assume both edges states to have the same Fermi velocity $v_F$. A difference in Fermi velocities could be absorbed into a redefinition of $p$.

At finite temperatures, similar steps as for filling factor $\nu=1$ lead to the following expression for the tunneling current
\begin{align}
\label{Current. Filling factor two. No interactions. Finite temperature}
& I_{dc}(V)/I_0=\frac{\xi T}{v_F}\sinh\left(\frac{V}{T}\right)\\
& \times \left[\mathcal{F}\left(0,\frac{V}{\pi T}\right)-\mathcal{N}(p,k \xi) \mathcal{F}\left(\frac{2\pi \xi T}{v_F}, \frac{V}{\pi T}\right)\right],\notag
\end{align}
with $\mathcal{F}(a,b)$ defined as in Eq.~(\ref{Dimensionless integral F(a,b)}). We have introduced the interference factor
\begin{equation}
\mathcal{N}(p,k \xi)=1-2p(1-p)[1-\cos(k \xi)],
\end{equation}
where $k=Bl$ is the momentum difference between the two edge channels when separated by a length $l$ in a magnetic field of strength $B$. This reflects the inherent relationship between momentum and position of QH edge states\cite{Patlatiuk}, where taking the difference avoids all dependence on the choice of gauge. The result for the zero-bias conductance at finite temperature reads
\begin{equation}
\frac{G(T)}{G_0}=1-\frac{2\pi \xi T\mathcal{N}(p,k \xi)}{v_F\sinh (2\pi \xi T/v_F)}.
\end{equation}
For $\xi T /v_F \ll 1$ we have $G(T)/G_0 \simeq 1-\mathcal{N}(p,k \xi)+(2 \mathcal{N}(p,k \xi)/3)(\pi \xi T/v_F)^2$. The leading order generally does not vanish and does not depend on temperature. Physically this is due to a circumvention of the Pauli blockade by allowing the electrons to tunnel simultaneously into different channels.

Employing Eq.~(\ref{Current. Filling factor two. No interactions. Finite temperature}), we get the result for the tunneling current at zero temperature
\begin{equation}
\label{zero temperature filling factor two no interactions}
I_{dc}(V)/I_0=\frac{2V \xi}{v_F}\left[1-\mathcal{N}(p,k \xi) \frac{\sin(2V\xi/v_F)}{2V \xi/v_F}\right].
\end{equation}
As before, the current vanishes if either $V \to 0$ or $\xi \to 0$. For tunneling into a single edge state ($p=1$ or $p=0$) one has $\mathcal{N} = 1$ and thus recovers the result from Eq.~(\ref{Current. Zero temperature. Filling factor one. DC regime}). It is interesting to note that for arbitrary $p$, one can still have $\mathcal{N} = 1$ if $\cos(k\xi) = 1$, which is likely due to destructive interference of the two tunneling events. While most of the parameters in the cosine argument are material specific, and thus hard to vary experimentally, it should be possible to observe this recovery of the Pauli blockade by varying the B-field within an interval maintaining the $\nu=2$ filling factor.

\begin{figure}
\includegraphics[width=\columnwidth]{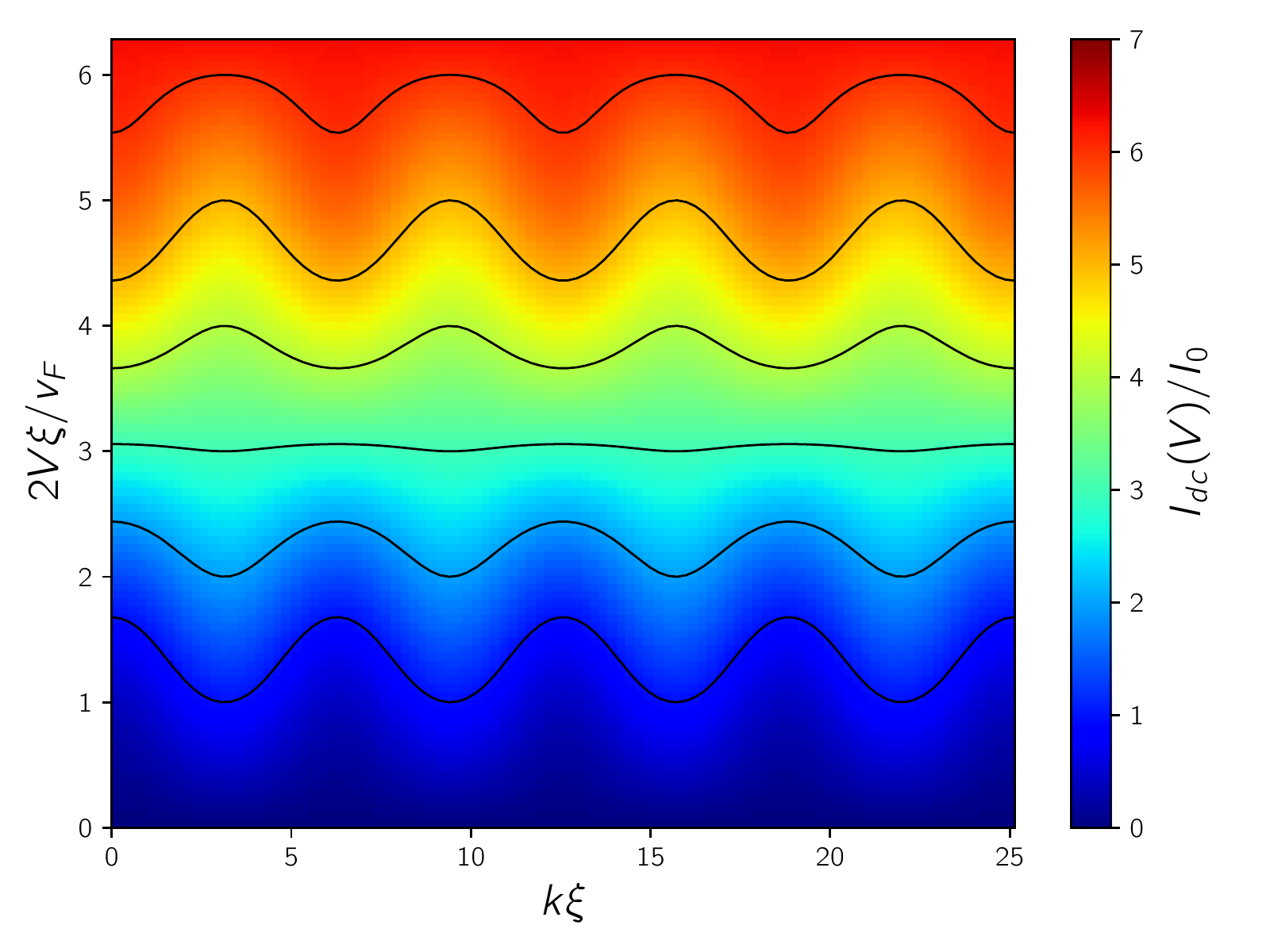}
\caption{\label{fig:three} At filling factor $\nu = 2$, the normalized tunneling current oscillates both with the dimensionless bias $2V\xi/v_F$ and the interference factor argument $k\xi$. These oscillations are shown at finite temperature, $2\xi T/v_F=0.5$, and tunneling probability $p=0.5$ (see Eq.~(\ref{Current. Filling factor two. No interactions. Finite temperature})). Black lines indicate integer values of the current.}
\end{figure}

In the limit $V\xi /v_F \ll 1$, we find that having two edge channels available and thus the possibility to avoid the Pauli blockade restores Ohmic behaviour: $I_{dc}(V)\propto (\tau \Delta/v_F)^2(1-\mathcal{N})V$, whereas the sub-leading term is proportional to $(\tau \Delta)^2 \mathcal{N} V^3 \xi^2 /v^4_F$. Eq.~(\ref{Current. Filling factor two. No interactions. Finite temperature}) is shown in Fig.~\ref{fig:three} for fixed finite temperature and tunneling probability. The oscillations with respect to the voltage are similar to those in Fig~\ref{fig:two}, while the oscillations with respect to the interference factor argument $k\xi$ are related to the Pauli blockade. For a fixed applied bias, these Pauli blockade oscillations are peaked at $k\xi=2 \pi n$, with $n \in \mathbb{N}$. At zero temperature these peaks become sharper, demonstrating a stronger blockade regime. Further, a straightforward calculation leads to the zero-temperature conductance $G(T=0)/G_0=1-\mathcal{N}(p,k \xi)$.

\begin{figure}
\includegraphics[width=\columnwidth]{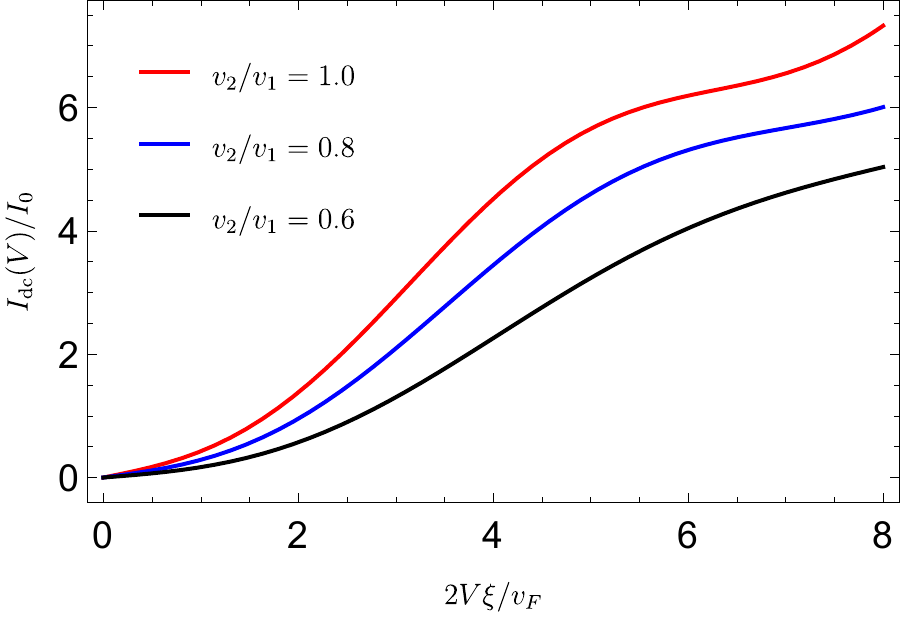}
\caption{\label{fig:four} When the electrons tunnel into two interacting QH edge states, we see a decrease in the magnitude of the normalized tunneling current, as well as longer oscillation periods with applied bias, for stronger interactions. The figure shows the case for finite temperature $2\xi T/v_F=0.5$ with an interaction parameter $v_2/v_1$ which has $v_2=v_F$ (see Eq.~(\ref{DC current. Filling factor two. Interactions. Finite temperature})), and where $v_2/v_1$ is the non-interacting case.}
\end{figure}

We now introduce electron-electron interactions both within a given edge state as well as between the two edge states. To study the effects of these interactions on the Pauli blockade, we start in the blockaded regime and thus assume that electrons only tunnel into one edge channel, corresponding to $p=1$ or $p=0$. To describe the edge states in the presence of interactions we use an effective field theory~\cite{Wen1, Wen2}. The edge state excitations are then described as collective fluctuations of the charge density $\hat{\rho}_j(x)=(1/2\pi)\partial_x \hat{\phi}_j(x)$, where the index $j=1,2$ labels the edge state and $\hat{\phi}_j(x)$ is a bosonic field operator which satisfies the standard commutation relations $[\hat{\phi}_i(x),\hat{\phi}_j(y)]=i\pi \delta_{ij}\text{sgn}(x-y)$. The Hamiltonian of the QH edge states is then given by
\begin{equation}
\label{QH Hamiltonian in presence of interactions}
\hat{H}_{QH}=\frac{1}{2} \sum_{ij=1,2} \int dx \int dy \hat{\rho}_i(x)V_{ij}(x,y) \hat{\rho}_j(y),
\end{equation}
where the interaction kernel is given by $V_{ij}(x,y)=(U+2\pi v_F \delta_{ij})\delta(x-y)$ with $U > 0$ describing the screened Coulomb interaction. The Hamiltonian $\hat{H}_{QH}$ can be diagonalized by the unitary transformation~\cite{Edvin1}
\begin{equation}
\label{Unitary transformation}
\hat{\phi}_{1}=\frac{1}{\sqrt{2}}(\hat{\chi}_1+\hat{\chi}_2), \quad \hat{\phi}_{2}=\frac{1}{\sqrt{2}}(\hat{\chi}_1-\hat{\chi}_2),
\end{equation}
which conserves the bosonic commutation relations $[\hat{\chi}_i(x),\hat{\chi}_j(y)]=i\pi \delta_{ij}\text{sgn}(x-y)$. Substituting these fields into the Hamiltonian~(\ref{QH Hamiltonian in presence of interactions}) we obtain
\begin{equation}
\label{QH hamiltonian in terms of charge and dipole mode}
\hat{H}_{QH}=\frac{1}{4\pi} \sum_{j=1,2} v_j \int dx  (\partial_x \hat{\chi}_j)^2,
\end{equation}
which now contains a fast charge mode ($j = 1)$ and a slow dipole mode ($j = 2$), with velocities $v_1=U/\pi + v_F$ and $v_2=v_F$, respectively. This bosonization procedure allows us to take into account electron-electron interactions with arbitrary strength explicitly and shows that the spectrum is split into two modes. Now, it is straightforward to calculate the four-point correlation functions using this diagonal Hamiltonian and the unitary transformation~(\ref{Unitary transformation}) (see App.~\ref{Sec:B}). Substituting the correlation functions from Eq.~(\ref{Four point correlation function}) into Eq.~(\ref{Tunneling current}), we get the following expression for the tunneling current at finite temperatures
\begin{align}
\label{DC current. Filling factor two. Interactions. Finite temperature}
& I_{dc}(V)/I_0=\frac{v_2}{v_1}\frac{2\xi T}{v_2}\sinh\left(\frac{V}{T}\right)\prod_{j=1,2}\sinh\left(\frac{\pi T\xi}{v_j}\right) \notag \\
& \times \mathcal{J}\left(\frac{2\pi T\xi}{v_1},\frac{2\pi T\xi}{v_2},\frac{2V}{\pi T}\right),
\end{align}
where the last factor has the integral form
\begin{equation}
\label{dimensional J(a1,a2,b) integral}
\mathcal{J}(a_1,a_2,b)=\int_{-\infty}^\infty dy \frac{\cosh^{-2}(y)\cos(by)}{\prod \limits_{i=1,2} \sqrt{\cosh(2y)+\cosh(a_i)}}.
\end{equation}
At low temperatures $T \xi/v_2 \ll 1$, the asymptotic form of the conductance is $G(T)/\mathcal{G}_0 \simeq (2/3)(\pi \xi T/\sqrt{v_1 v_2})^2 $, where $\mathcal{G}_0=2(\tau \Delta)^2/\pi v_1 v_2$ is a normalization coefficient. At zero temperature and $V \xi /v_2 \ll 1$ with $v_1 > v_2$ we get $I_{dc}(V) \propto (2/3\pi)(V^3 \xi^2)/(v_1 v_2)^2$, resulting in vanishing zero-bias conductance. Thus we get the same result as in case of filling factor $\nu=1$ in Eq.~(\ref{Current. Zero temperature. Filling factor one. DC regime}) at $V \xi /v_F \ll 1$. The Pauli blockade persists even with cross-channel interaction. As one can see, the interaction renormalizes the Fermi velocity, so to obtain the current at $T \rightarrow 0$ one has to change $v_F$ to $\sqrt{v_1 v_2}$ in asymptotics of Eq.~(\ref{Current. Zero temperature. Filling factor one. DC regime}). The dependence of the tunneling current on the applied bias at different interaction parameters, $v_2/v_1$, is shown in Fig.~\ref{fig:four}. Here $v_2/v_1=1$ corresponds to the non-interacting case. One can see that the interaction parameter slightly decreases the magnitude of the tunneling current in comparison with the non-interacting regime, while the oscillation period is increased.

\subsection{Filling factor $\nu=1/(2n+1)$}
\label{Filling factor 1/3}
The fractional QH edge state with Laughlin filling
factor $\nu = 1/(2n + 1), n \in \mathbb{N}$, consists of a single channel with a free bosonic field $\hat{\phi}(x)$ propagating with velocity $v$. The electron operator is given by the vertex operator $\hat{\psi}(x) \propto e^{i \hat{\phi}(x)/\sqrt{\nu}}$ \cite{Wen1, Wen2}. We can then repeat the steps of the previous sections to get the tunneling current at finite temperature
\begin{align}
\label{DC current. Filling factor 1/3. Finite temperature}
& I_{dc}(V)/\tilde{I}_0=2^{1/\nu} \frac{\xi T}{v}\left(\frac{rT}{2v}\right)^{2/\nu-2} \notag \\
& \times \sinh^{2/\nu}\left(\frac{\pi \xi T}{v}\right)\sinh\left(\frac{V}{T}\right) \mathcal{Q}\left(\frac{2\pi \xi T}{v},\frac{2V}{\pi T}\right),
\end{align}
where $r$ is an ultraviolet cut-off, $\tilde{I}_0=(\tau \Delta)^2/\pi v \xi$ is the normalization coefficient, and we use the dimensionless integral
\begin{equation}
\label{Dimensionless integral Q(a,b)}
\mathcal{Q}(a,b)= \int_{-\infty}^\infty dy \frac{\cosh^{-2/\nu}(y) \cos(by)}{[\cosh(2y)+\cosh(a)]^{1/\nu}}.
\end{equation}
At $ T \xi/v \ll 1$ we get the asymptotic behavior of the conductance
\begin{equation}
\label{Conductance. Filling factor 1/3. Finite temperature. Asymptotics}
 \frac{G(T)}{\tilde{G}_0} \simeq \frac{\sqrt{\pi}}{2}\frac{\Gamma(2/\nu)}{\Gamma(1/2+2/\nu)} \left(\frac{rT}{2v}\right)^{2/\nu-2}\left(\frac{\pi \xi T}{v}\right)^{2/\nu},
\end{equation}
where $\Gamma(x)$ denotes the gamma function and $\tilde{G}_0=2(\tau \Delta)^2/\pi v^2$. Further, at zero temperature and low voltages $V \xi/v \ll 1 $, using Eq.~(\ref{DC current. Filling factor 1/3. Finite temperature}), we find that the current has the form
\begin{equation}
\label{DC current. Filling factor 1/3. Zero temperature. Asymptotics}
\begin{split}
& I_{dc}(V)/\tilde{I}_0 \simeq \frac{2\pi^2 v \xi}{r^2}\left(\frac{2 r \xi}{\pi v^2}\right)^{2/\nu} \frac{V^{4/\nu -2}}{\Gamma(4/\nu)}.
\end{split}
\end{equation}
Consequently, the conductance vanishes as in the case of filling factor $\nu=1$, i.e. $G=0$. This result can also be obtained from Eq.~(\ref{Conductance. Filling factor 1/3. Finite temperature. Asymptotics}) at $T \to 0$. This is related to the power-law behavior of the tunneling current with respect to the applied voltage due to the positive integer power in Eq.~(\ref{DC current. Filling factor 1/3. Zero temperature. Asymptotics}). Even though we have tunneling between two effectively bosonic systems, the Pauli blockade persists and makes the QPC an insulator at zero bias. At filling factor $\nu=1$ the results of this subsection coincide with the results of subsection~(\ref{Filling factor one}). 

\section{Finite-frequency noise in the DC regime}
\label{Sec:IV}
In this section, we consider the finite-frequency noise in the case of an applied dc voltage. The exact experimentally measurable current noise depends on the details of the setup, so we calculate the non-symmetrized current correlation function from which other forms of noise, e.g., the symmetrized noise, can be obtained~\cite{Lesovik, Martin1, Zamoum, Edvin2}. It is defined as
\begin{equation}
\label{Definition of finite frequency noise in presence of dc bias}
S_{dc}(\omega, V)=\int_{-\infty}^{+\infty}  dt e^{i \omega t}  \langle \delta \hat{J}(t)\delta \hat{J}(0) \rangle,
\end{equation}
where $\delta\hat{J}(t)=\hat{J}(t)-\langle \hat{J}(t)\rangle$ and the average is taken with respect to the dc biased ground state of QH system and superconductor. Using the time translation invariance of the vertex operators (see App.~\ref{Sec:C}), the noise can be written to the lowest order of the tunneling coupling as
\begin{equation}
S_{dc}(\omega, V)= g(\omega+2V)+g(\omega-2V),
\end{equation}
where the correlation function on the right is given by
\begin{equation}
\label{Expression for finite frequency noise in presence of dc bias. Plus and minus terms}
g(\omega)=4(\tau \Delta)^2 \int_{-\infty}^{+\infty} dt e^{i \omega t}\langle \hat{A}^{\dagger}(t)\hat{A}(0)\rangle_0.
\end{equation}
It is worth pointing out that the shot noise at $\omega=0$ is determined by the anti-commutator of $\hat{A}$ operators, in contrast to the tunneling current in Eq.~(\ref{Tunneling current}), namely~\cite{Edvin2} $S_{dc}(0,V)=4(\tau \Delta)^2 \int_{-\infty}^{+\infty} dt e^{2iVt}\langle \{A^{\dagger}(t),A(0)\} \rangle_0 $. The noise can be symmetrized as the even combination of the two non-symmetrized terms, $[S(\omega)+S(-\omega)]/2$, and whether measuring the non-symmetrized or the symmetrized noise is possible depends on the experimental detector~\cite{Lesovik}.

\subsection{Filling factor $\nu=2$}
We start again by considering a system with positive bias voltage $V>0$, no interactions and Cooper pairs tunneling simultaneously into both edge channels (see Sec.~\ref{Filling factor two. DC current}). Using Eq.~(\ref{Definition of finite frequency noise in presence of dc bias}) and the two-point correlation functions from App.~\ref{Sec:A}, the noise at finite temperature becomes
\begin{align}
\label{Noise finite temperature. Filling factor two. No interactions}
& S_{dc}(\omega, V)/I_0=\sum_{\sigma=\pm} \frac{\xi T}{v_F}\exp\left(\frac{\omega + 2\sigma V}{2T}\right) \\
& \times \left[\mathcal{F}\left(0,\frac{\omega + 2\sigma V}{2 \pi T}\right)-\mathcal{N}(p,k\xi)  \mathcal{F}\left(\frac{2\pi \xi T}{v_F},\frac{\omega + 2\sigma V}{2\pi T}\right)\right] \notag,
\end{align}
where $\mathcal{F}(a,b)$ is defined in Eq.~(\ref{Dimensionless integral F(a,b)}) and the normalization coefficient $I_0$ is given after Eq.~(\ref{Tunneling current. Finite temperature. Filling factor one. DC regime}).

\begin{figure}
\includegraphics[width=\columnwidth]{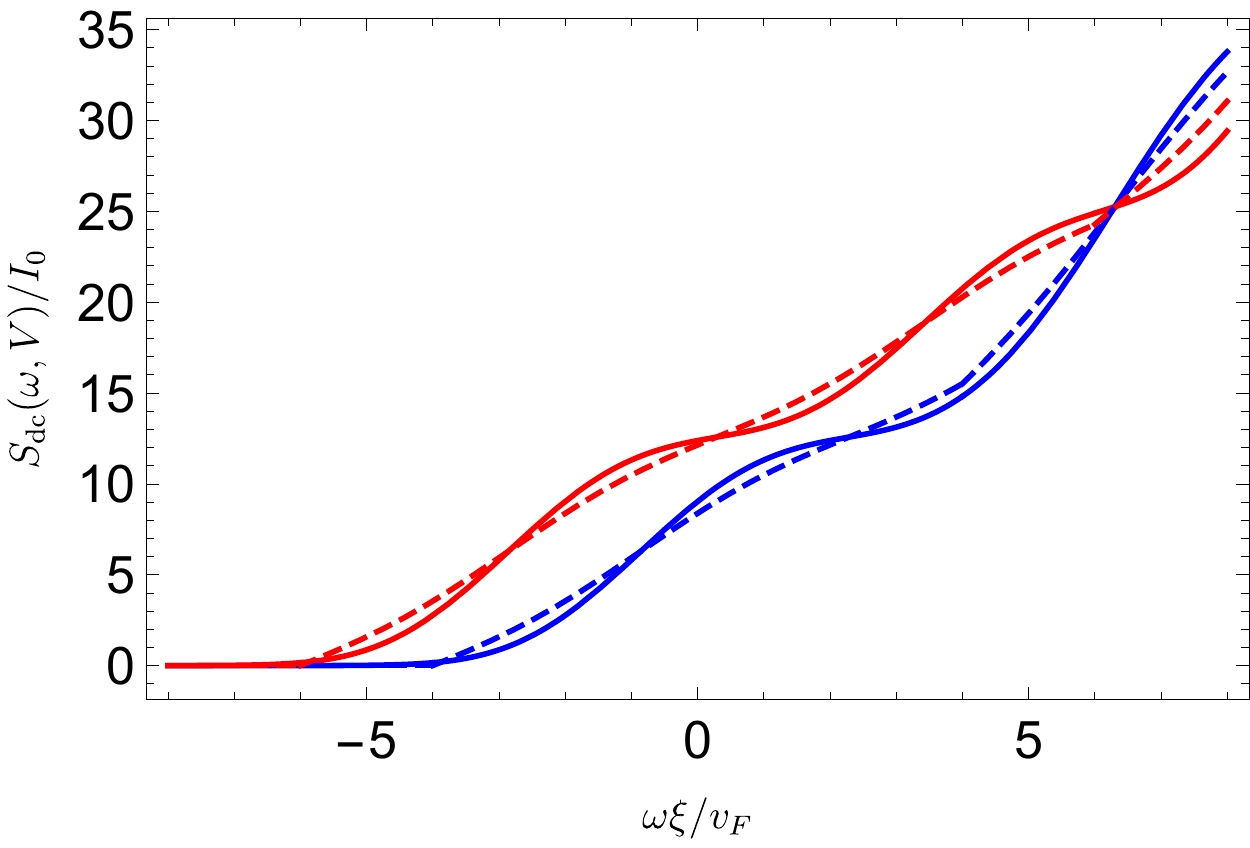}
\caption{\label{fig:five} The normalized finite frequency non-symmetrized noise of the dc tunneling current $S_{dc}(\omega, V)/I_0$ at filling factor $\nu = 2$ as a function of the dimensionless frequency $\omega \xi/v_F$ at applied voltages $2V \xi/v_F=4$ (blue lines) and $2V\xi /v_F=6$ (red lines). Here, the solid lines correspond to finite temperature $2 T\xi/v_F=0.5$ and the dashed lines correspond to zero temperature. Notably, we find no resonances (singularities). We have $p=0.5$ and $k\xi=2\pi/3$. See Eqs.~(\ref{Noise finite temperature. Filling factor two. No interactions}) and (\ref{Noise zero temperature. Filling factor two. No interactions}).}
\end{figure}

At zero temperature we can use Eq.~(\ref{Noise finite temperature. Filling factor two. No interactions}) to obtain the expression
\begin{align}
\label{Noise zero temperature. Filling factor two. No interactions}
& S_{dc}(\omega, V)/I_0= \sum_{\sigma=\pm} 2 \theta (\omega \xi/v_F+2\sigma V \xi/v_F) \\
& \times \left\{\frac{|\omega+2\sigma V|\xi}{v_F}-\mathcal{N}(p,k\xi)\sin\left(\frac{|\omega+2\sigma V|\xi}{v_F}\right)\right\}, \notag
\end{align}
where $\mathcal{N}(p,k\xi)$ is given in Eq.~(\ref{zero temperature filling factor two no interactions}) and $\theta(x)$ is the Heaviside step function. The dependence of noise on frequency at zero and finite temperatures is shown in Fig.~(\ref{fig:five}). The oscillations are again related to the tunneling of electrons into two spatially separated points, $x=0$ and $x=\xi$. At small frequencies $0<\omega \ll 2V$, the linear frequency dependent part appears in a sub-leading term, namely $S_{dc}(\omega, V)/I_0 \approx 2(2V \xi/v_F-\mathcal{N}\sin(2V\xi/v_F))+2(1-\mathcal{N}\cos(2V \xi/v_F))(\omega \xi/v_F)$. At large frequencies $\omega \gg 2V>0$, the frequency dependent part appears in the leading order, $S_{dc}(\omega,V)/I_0 \approx 4\omega \xi/v_F$. Furthermore, we calculate the derivative of the shot noise with respect to the applied bias at $V \to 0$ and get
\begin{equation}
\label{Differential shot noise. Filling factor two, no interaction. Zero temperature}
G^{-1}_0 \left. \frac{\partial S_{dc}(0,V)}{\partial V}\right|_{V \to 0}=2[1-\mathcal{N}(p,k\xi)],
\end{equation}
where the prefactor $G_0$ on the left is given after Eq.~(\ref{Conductance. Zero temperature. Filling factor one}). Here, as with the corresponding conductance, we see that at filling factor $\nu=2$ the Pauli blockade is lifted. At low temperatures $\xi T/v_F \ll 1$, the sub-leading correction to Eq.~(\ref{Differential shot noise. Filling factor two, no interaction. Zero temperature}) is given by $(4/3)\mathcal{N}(p,k\xi)(\pi \xi T/v_F)^2$, i.e. the temperature independence is only to leading order.

Furthermore, using Eqs.~(\ref{Current. Filling factor two. No interactions. Finite temperature}) and (\ref{Noise finite temperature. Filling factor two. No interactions}), and $\mathcal{F}(a,-b)=\mathcal{F}(a,b)$, one can see that the Fano factor has the well-known form~\cite{Glattli1}
\begin{equation}
\label{Fano factor at finite temperature}
S_{dc}(0,V)/I_{dc}(V)=2\coth\left(\frac{V}{T}\right).
\end{equation}
Therefore, the current fluctuations satisfy a classical Poissonian shot noise form~\cite{Glattli1}. At zero temperature, we have $\coth(V/T) \to 1$, so the Fano factor becomes $S_{dc}(0,V)/I_{dc}(V)=2$, which can be taken as an indication that the elementary charge carriers tunneling through the QPC are indeed charge-$2e$ Cooper pairs.

The result for filling factor $\nu=1$ can be obtained setting $\mathcal{N}(p,k\xi)=1$ at $p=1$ or $p=0$. In particular, at low temperatures and bias voltage, $\xi T/v_F \ll 1$ and $V/T \ll 1$, we get
\begin{equation}
\label{Shot noise. Finite temperature. Filling factor one. DC regime}
S_{dc}(0,V)/I_0 \simeq \frac{8 V \xi}{3v_F}\left(\frac{\pi \xi T}{v_F}\right)^2.
\end{equation}
Thus, the right hand side of Eq.~(\ref{Shot noise. Finite temperature. Filling factor one. DC regime}) becomes proportional to the differential conductance (\ref{Conductance. Zero temperature. Filling factor one}) at $V \to 0$, i.e. $\partial S_{dc}(0,V)/\partial V \propto (\xi T/v_F)^2$.

Taking into account electron-electron interactions, assuming tunneling into only one channel, and repeating the steps leading to Eq.~(\ref{DC current. Filling factor two. Interactions. Finite temperature}) and ~(\ref{Noise finite temperature. Filling factor two. No interactions}) for finite temperatures yields the following result
\begin{equation}
\label{Finite frequency noise. Filling factor two with interactions. Finite temperature}
\begin{split}
& S_{dc}(\omega, V)/I_0=\sum_{\sigma=\pm} \frac{v_2}{v_1}\frac{2 \xi T}{v_2}\exp\left(\frac{\omega + 2\sigma V}{2T}\right) \\
& \times \prod_{i=1,2} \sinh\left(\frac{\pi T \xi}{v_i}\right) \mathcal{J}\left(\frac{2\pi T \xi}{v_1}, \frac{2 \pi T \xi}{v_2}, \frac{\omega + 2\sigma V}{\pi T}\right),
\end{split}
\end{equation}
where $I_0$ is given in Eq.~(\ref{DC current. Filling factor two. Interactions. Finite temperature}), and $\mathcal{J}$ is given in Eq.~(\ref{dimensional J(a1,a2,b) integral}). Calculating the zero-frequency noise, one finds that the Fano factor is given by Eq.~(\ref{Fano factor at finite temperature}), as it is expected. It is worth mentioning here that at $V \to 0$, the differential shot noise vanishes.

At zero temperature, using Eq.~(\ref{Finite frequency noise. Filling factor two with interactions. Finite temperature}), for the shot noise at $\omega=0$ and $V\xi/v_1, V \xi /v_2 \ll 1$ we get
\begin{equation}
\label{Shot noise. Filling factor two with interactions. Asymptotics}
S_{dc}(0,V)/I_0 \simeq \frac{8 v^2_2}{3v^2_1}\frac{V^3 \xi^3}{v^3_2}.
\end{equation}
Direct calculations using the asymptotic result in Eq.~(\ref{Shot noise. Filling factor two with interactions. Asymptotics}) or the exact expression at $\omega=0$ in Eq.~(\ref{Finite frequency noise. Filling factor two with interactions. Finite temperature}), give that $\partial S_{dc}(0,V)/ \partial V=0$ at $V \to 0$, which is caused by the Pauli blockade.

\subsection{Filling factor $\nu=1/(2n+1)$}
Repeating the steps of the previous subsections, in case of finite temperature we get the following result for noise at fractional filling factors
\begin{equation}
\label{Noise. Fractional filling factor. Finite temperature}
\begin{split}
& S_{dc}(\omega, V)/\tilde{I}_0=\sum_{\sigma=\pm} 2^{1/\nu}\frac{\xi T}{v}\left(\frac{rT}{2v}\right)^{2/\nu -2} \sinh^{2/\nu}\left(\frac{\pi \xi T}{v}\right) \\
& \times \exp\left(\frac{\omega + 2\sigma V}{2T}\right)\times \mathcal{Q}\left(\frac{2\pi \xi T}{v}, \frac{\omega + 2\sigma V}{\pi T}\right),
\end{split}
\end{equation}
where $\mathcal{Q}(a,b)$ is given in Eq.~(\ref{Dimensionless integral Q(a,b)}) and the normalization factor is given in Eq.~(\ref{DC current. Filling factor 1/3. Finite temperature}). Taking this equation at $T \xi /v \ll 1$ one can show that the differential shot noise at zero bias is proportional to the conductance, namely $\partial S_{dc}(0,V)/ \partial V\propto (rT/2v)^{2/\nu-2}(\pi \xi T/v)^{2/\nu}$. This expression vanishes at $T \to 0$. In particular we find that at zero temperature, zero frequency, $\omega=0$, and $V\xi/v \ll 1$ we obtain the asymptotic expression for shot noise
\begin{equation}
\label{Shot noise. Zero temperature. Filling factor 1/3. Asymptotics}
S_{dc}(0,V)/\tilde{I}_0 \simeq \frac{4\pi^2}{\Gamma[4/\nu]} \frac{v \xi}{r^2 V}\left(\frac{2r \xi V^2}{\pi v^2}\right)^{2/\nu}.
\end{equation}
This expression results in vanishing differential shot noise, namely $\partial S_{dc}(0,V)/\partial V=0$ at $V \to 0$. It is worth mentioning that at $\nu=1$ the result of this subsection agrees with the expressions of the previous subsection and Eq.~(\ref{Fano factor at finite temperature}) for Fano factor at finite temperatures is satisfied. We note that no resonances (singularities) appear in Eqs.~(\ref{Noise zero temperature. Filling factor two. No interactions}), (\ref{Finite frequency noise. Filling factor two with interactions. Finite temperature}) or (\ref{Noise. Fractional filling factor. Finite temperature}). For instance, at zero temperature, this can be seen from the fact that the power-law correlation functions result in the linear frequency behavior of noise $S(\omega, V) \propto \omega$ at small, $\omega \ll 2V$, and large, $\omega \gg 2V$, frequencies.

\section{Tunneling current and finite frequency noise in the AC regime}
\label{Sec:V}
To study the case of time-dependent voltage, we assume a periodic bias of the form $\tilde{V}(t)=V+V_1\cos(\Omega t)$, where $\Omega$ is the driving frequency. The time-dependent part of such bias averages to zero over one period $\mathcal{T}=2\pi/ \Omega$, and the dc part of the time averaged tunneling current in the case of ac bias is given  by
\begin{equation}
I=\frac{2}{\mathcal{T}} \int^{\mathcal{T}}_0 dt \int^t_{-\infty} dt^{\prime} \text{Re}\left\{e^{i\int^t_{t^{\prime}} \tilde{V}(t^{\prime})} \langle [\hat{A}^{\dagger}(t),\hat{A}(t^{\prime})]\rangle_0 \right\}.
\end{equation}
With the exact form of the vertex operators from App.~\ref{Sec:B} and App.~\ref{Sec:C}, and using an expansion in terms of Bessel functions, $\exp[i \lambda \sin \varphi]=\sum^{\infty}_{n=-\infty}J_n(\lambda)\exp[i n \varphi ]$, we find
\begin{equation}
\label{tunneling current with ac bias. Result}
I=\sum^{+\infty}_{n=-\infty} J^2_n(2V_1 /\Omega)\  I_{dc}(V+n\Omega/2),
\end{equation}
where $I_{dc}(V+n\Omega/2)$ has been calculated in Eqs.~(\ref{Tunneling current. Finite temperature. Filling factor one. DC regime}), (\ref{Current. Filling factor two. No interactions. Finite temperature}), (\ref{DC current. Filling factor two. Interactions. Finite temperature}) and (\ref{DC current. Filling factor 1/3. Finite temperature}). At $\Omega \to 0$ and $V_1 \to 0$, the sum of Floquet factors goes to one, i.e $\sum^{\infty}_{n=-\infty}J^2_n(2V_1/\Omega) \to 1$ and thus we recover the result of Eq.~(\ref{Tunneling current}) in the case of dc bias for all filling factors.

\begin{figure}
\includegraphics[width=\columnwidth]{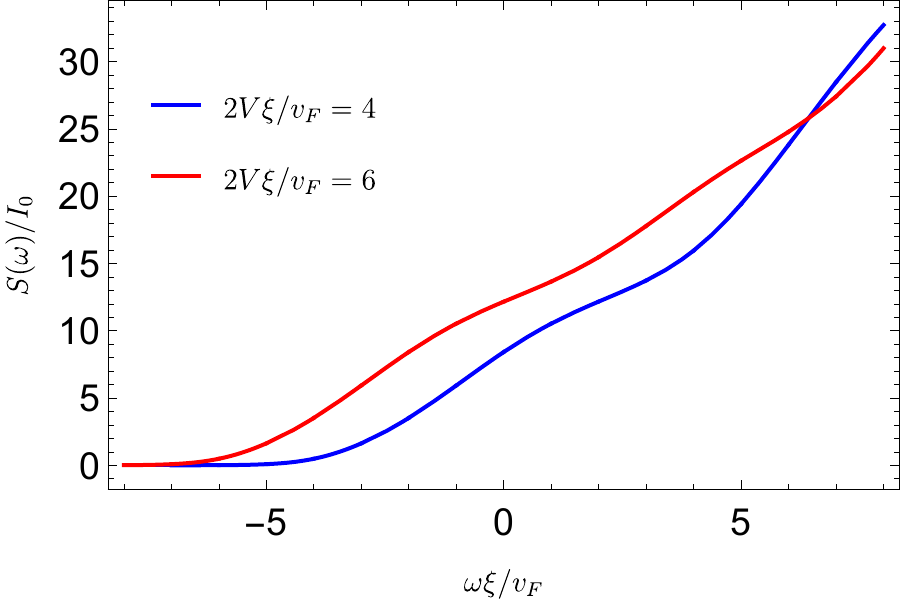}
\caption{\label{fig:six} The normalized finite frequency non-symmetrized noise of the tunneling current in presence of periodic time-dependent (ac) bias, $\tilde{V}(t)=V+V_1 \cos(\Omega t)$ at zero temperature for different $2V \xi/v_F$, see Eq.~(\ref{Finite frequency noise. AC regime}). Again we find no resonances. In this figure we show the case of filling factor $\nu=1$ and choose $S_{dc}(\omega, V)$ from Eq.~(\ref{Noise zero temperature. Filling factor two. No interactions}). We set $p=0$ or $1$, $2V_1 \xi/v_F=2$ and $\Omega \xi/v_F=1$.}
\end{figure}

We proceed by calculating the noise in the presence of such an ac bias voltage. Again, we consider the finite-frequency noise averaged over a drive period $\mathcal{T}$. Due to the drive, this can be regarded as noise due to photon assisted electron transport across the QPC. The time-averaged photon assisted finite-frequency noise is given by the following Wigner transformation
\begin{equation}
S(\omega)=\frac{1}{\mathcal{T}} \int^{\mathcal{T}}_0 d\tau \int_{-\infty}^{+\infty} d\tau^{\prime}S(\tau+\tau^{\prime}/2, \tau-\tau^{\prime}/2)e^{i\omega \tau^{\prime}},
\end{equation}
where we have introduced the ``center of mass'' and ``relative'' time variables, $\tau=(t+t^{\prime})/2$ and $\tau^{\prime}=t-t^{\prime}$, respectively. The integrand includes the current-current correlation function $S(t,t^{\prime})=\langle \delta \hat{J}(t) \delta \hat{J}(t^{\prime}) \rangle$ with $\delta \hat{J}(t)=\hat{J}(t)-\langle \hat{J}(t)\rangle$ and the average is performed with respect to the biased ground state of the system. Using again an expansion of the exponent in terms of Bessel functions, the time invariance of the vertex correlation functions (see App.~\ref{Sec:C}) and $J^2_{-n}(x)=J^2_n(x)$ we get the final result for finite-frequency noise
\begin{equation}
\label{Finite frequency noise. AC regime}
S(\omega)=\sum^{\infty}_{n=-\infty}J^2_n (2V_1/\Omega) \ S_{dc}(\omega, V+n\Omega/2),
\end{equation}
where $S_{dc}(\omega,V+n\Omega/2)$ has been calculated in Eqs.~(\ref{Noise finite temperature. Filling factor two. No interactions}), (\ref{Noise zero temperature. Filling factor two. No interactions}), (\ref{Finite frequency noise. Filling factor two with interactions. Finite temperature}) and (\ref{Noise. Fractional filling factor. Finite temperature}). Here again, as $\Omega \to 0$ and $V_1 \to 0$, the sum of Floquet factors goes to one, i.e $\sum^{\infty}_{n=-\infty}J^2_n(2V_1/\Omega) \to 1$ and thus we recover the result of Eq.~(\ref{Definition of finite frequency noise in presence of dc bias}) in the case of dc bias. The dependence of noise on frequency, Eq.~(\ref{Finite frequency noise. AC regime}) at filling factor $\nu=1$ is presented on Fig.~\ref{fig:six}. As in case of dc bias, there are no resonances (singularities) in the frequency dependent noise, which is a consequence of the positive power-law behaviour with respect to frequency (see the discussion in the last paragraph of Sec.~\ref{Sec:IV}.) In Ref.~[\onlinecite{Glattli3}], in the low frequency regime, it was experimentally demonstrated that a similar equation, $\Lambda=\sum^{n=\infty}_{n=-\infty}J^2_n(V_1/\Omega)\Lambda_{dc}(V+n\Omega)$, holds for electron current, heat current and shot noise under ac bias over a QPC contact between two edge states. According to Eq.~(\ref{Finite frequency noise. AC regime}), this statement holds in one more general case, namely for finite-frequency noise. This result can be used to interpret the experiments on dynamical response of Laughlin anyons in presence of time-dependent bias~\cite{Glattli3}.

\section{Conclusion}
\label{Sec:VI}
In this paper we have studied tunneling between a superconductor and a QH edge state at different filling factors, namely $\nu=1$, $\nu=2$ and $\nu=1/(2n+1)$. To account for electron-electron interaction in the QH edge state, we used a low-energy effective theory based on bosonization. In the bosonic picture of collective excitations, the spectrum splits into two modes, namely the fast charge mode and slow dipole mode. Exact diagonalization allows us to calculate the two- and four-point equilibrium correlation functions, which are necessary to evaluate the transport properties of system, such as current and noise. We investigated the tunneling between the QH edge states and the superconductor to the lowest order in the tunneling coupling under the dc and ac biases.

For filling factor $\nu=1$, at zero temperature and $V \xi/v_F \ll 1$, we found that the tunneling current is proportional to $I_{dc}(V) \propto (\tau \Delta)^2 \xi^2 V^3/v^4_F$, which is a manifestation of non-Ohmic behavior. This scaling of the tunneling current with the applied dc bias results in a vanishing conductance. At finite temperatures, at $\xi T/v_F \ll 1$ and $V/T \ll 1$, the current is proportional to the applied bias, and the density of states is renormalized by the dimensionless factor $\xi T/v_F$, namely $I_{dc}(V) \propto (\tau \Delta /v_F)^2 (\xi T/v_F)^2 V$. In addition to the tunneling current, we presented results for the finite-frequency current noise. The ratio between shot noise and tunneling current, known as the Fano factor, was found to be $S_{dc}(0,V)/I_{dc}(V)=2$. Thus, the differential shot noise, $\partial S_{dc}(0,V)/\partial V$ at $V \to 0$ vanishes as well. At finite temperatures the Fano factor has the form $S_{dc}(0,V)/I_{dc}(V)=2\coth(V/T)$. As a result, at $\xi T/v_F \ll 1$, the differential shot noise at $V \to 0$ is proportional to the conductance and vanishes as $T \to 0$.

For filling factor $\nu=2$, in case of simultaneous tunneling of a Cooper pair into different QH channels, the situation changes drastically. At zero temperature and $V \xi /v_F \ll 1$, the current manifests Ohmic behavior to leading order, $I_{dc}(V) \propto (\tau \Delta/v_F)^2 (1-\mathcal{N})V$, where $0 < \mathcal{N} < 1$ in case of simultaneous tunneling into two edge states. The shot noise is proportional to the current, $S_{dc}(0,V)/I_{dc}(V)=2$, so the differential shot noise at $V \to 0$ is generally not equal to zero. At low temperatures $\xi T/v_F\ll 1$ the leading behavior of both the conductance and the differential shot noise does not depend on temperature. The temperature dependence appears only in a subleading correction proportional to $\mathcal{N}(\xi T/v_F)^2$. In the presence of electron-electron interaction the results are qualitatively similar, but one has to replace the Fermi velocity by the geometric average of the velocities of the charged and dipole modes, $\sqrt{v_1 v_2}$.

For filling factor $\nu=1/(2n+1)$, the power-law behavior of transport quantities depends on $\nu$. At $V \xi /v \ll 1$, the current is given by $I_{dc}(V) \propto V^{4/\nu -1}$ and the conductance vanishes. At low temperatures, $\xi T/v \ll 1$, we have $I_{dc}(V) \propto T^{4/\nu -2} V$ and the conductance depends on temperature. The behavior of the differential shot noise at $V \to 0$ with respect to temperature is identical to that of the conductance, namely $\propto T^{4/\nu-2}$.

We also provided a general expression for the tunneling current and the finite-frequency noise in the presence of a periodic ac bias voltage. This result, valid for all filling factors considered, demonstrates that the current and finite-frequency noise can be expressed as the sum of dc currents and noise terms with Floquet coefficients. Recently, it was experimentally found that an expression similar to our result (\ref{Finite frequency noise. AC regime}) holds for shot noise \cite{Glattli3}. We have found that this statement holds in the more general case of finite-frequency noise.

At Laughlin filling factors, in addition to the Coulomb blockade~\cite{Kane3}, it has been found that the vanishing conductance~\cite{Fisher} and differential shot noise at low temperatures is a consequence of an additional suppression mechanism called the Pauli blockade: after the tunneling of the first electron of a Cooper pair the tunneling of the second electron into the QH edge state is suppressed up to times $\xi/v_F$, where $v_F$ is the velocity of the edge excitations, due to the Pauli exclusion principle. At filling factor $\nu=2$, in the case of simultaneous tunneling of a Cooper pair into both channels, the Pauli blockade is partially removed. Electron-electron interactions do not change the physics qualitatively but result in a renormalization of the Fermi velocity. Finally, as a future perspective, it would be interesting to consider a similar problem in the context of levitonic physics~\cite{Glattli2}, where the injection of single particles due to tailored voltage pulses is investigated.

\acknowledgments
We are grateful to Christian Glattli for fruitful discussions. The authors acknowledge financial support from the National Research Fund Luxembourg under Grants CORE C19/MS/13579612/HYBMES and ATTRACT 7556175.

\appendix
\begin{widetext}
\section{Two-point correlation function}
\label{Sec:A}
In this Appendix we calculate the two-point correlation function of right-moving fermions $G_j(x,t;x^{\prime},t^{\prime})$ at filling factor $\nu=1$ and at finite temperature $T$. We use bosonization technique, which is further necessary to take into account the electron-electron interaction. Here the subscript $j=1,2$ denotes the QH channel. According to bosonization technique, we can write
\begin{equation}
G_j(x,x^{\prime};t,t^{\prime})=\langle \hat{\psi}^{\dagger}_j(x,t)\hat{\psi}_j(x^{\prime},t^{\prime})\rangle_0=\frac{1}{r}\langle e^{-i\hat{\phi}_j(x,t)}e^{i\hat{\phi}_j(x^{\prime},t^{\prime})} \rangle_0=\frac{1}{r}e^{M(x,t;x^{\prime}, t^{\prime})}
\end{equation}
where $r$ is an ultra-violet cut-off, and in Gaussian approximation under consideration the exponent is given by
\begin{equation}
M(x,t;x^{\prime}, t^{\prime})=-\frac{1}{2}\langle \hat{\phi}^2_j(x,t)\rangle_0 -\frac{1}{2}\langle \hat{\phi}^2_j(x^{\prime},t^{\prime})\rangle_0 + \langle \hat{\phi}_j(x,t)\hat{\phi}_j(x^{\prime},t^{\prime})\rangle_0.
\end{equation}
Using the expansion of bosonic field in terms of creation and annihilation operators of bosons, we get the following expression (zero modes are ignored)
\begin{equation}
M(x,t;x^{\prime},t^{\prime})=\int^{\infty}_0 \frac{dk}{k}e^{-\frac{rk}{2\pi}}\left[\left(1+f_B(k)\right)\left(e^{ik[X(t)-X^{\prime}(t^{\prime})]}-1\right)+f_B(k)\left(e^{-ik[X(t)-X^{\prime}(t^{\prime})]}-1\right)\right],
\end{equation}
where $X=x-v_jt$, $X^{\prime}=x^{\prime}-v_jt^{\prime}$ and $f_B(k)=(e^{v_jk/T}-1)^{-1}=\sum^{\infty}_{n=1}e^{-v_j \beta k n}$ is an equilibrium bosonic distribution function with inverse temperature $\beta=1/T$. Further integration with respect to momentum variable $k$ gives
\begin{equation}
M(x,t;x^{\prime},t^{\prime})=\log\left[\frac{ir/2\pi}{X(t)-X^{\prime}(t^{\prime})+ir/2\pi}\right]-\sum^{\infty}_{n=1}\log\left[1+\frac{[\pi(X(t)-X^{\prime}(t^{\prime})+ir/2\pi)/v_j\beta]^2}{\pi^2 n^2}\right].
\end{equation}
Next, exponentiating the above relation and using the definition of hyperbolic sine
\begin{equation}
\sinh(z)=z\prod^{\infty}_{n=1}\left(1+\frac{z^2}{\pi^2 n^2}\right),
\end{equation}
we finally get the result for two-point correlation function at finite temperature
\begin{equation}
\label{Finite temperature. Two-point correlation function. Section A}
G_j(x,x^{\prime};t,t^{\prime})=\frac{-iT}{2v_j}\frac{1}{\sinh\left[\pi T\left(t-t^{\prime}-(x-x^{\prime})/v_j-i\gamma\right)\right]}, \quad \gamma \to +0.
\end{equation}
The correlation function at zero temperature is obtained, using that $\sinh(x) \sim x$, namely
\begin{equation}
\label{Zero temperature. Two-point correlation function. Section A}
G_j(x,x^{\prime};t,t^{\prime})=\frac{-i}{2\pi v_j}\frac{1}{t-t^{\prime}-(x-x^{\prime})/v_j-i\gamma}.
\end{equation}

\section{Four-point correlation function}
\label{Sec:B}
In this Appendix we derive the expression for four-point correlation function~\cite{Edvin4}. We again use the Gaussian character of theory to calculate it, namely the average of four vertex operators is written as the exponent of combination of averages of bosonic field. To demonstrate this, we use bosonization technique to rewrite the four-point correlation function, namely
\begin{equation}
\mathcal{L}_1=\langle \hat{\psi}^{\dagger}_1(x_1,t_1) \hat{\psi}^{\dagger}_1(x_2,t_2) \hat{\psi}_1(x_3,t_3) \hat{\psi}_1(x_4,t_4)\rangle_0=\frac{1}{r^2}\langle e^{-i\hat{\phi}_1(x_1,t_1)} e^{-i\hat{\phi}_1(x_2,t_2)} e^{i\hat{\phi}_1(x_3,t_3)} e^{i\hat{\phi}_1(x_4,t_4)}\rangle_0,
\end{equation}
where we have omitted the arguments of $\mathcal{L}_1$ and the average is taken with respect to equilibrium zero density matrix, $\hat{\rho}_0$. Next, using the Eq.~(\ref{Unitary transformation}) from main text, the above expression can be rewritten as a product of two four-point vertex correlation functions corresponding to charged and dipole modes in presence of interaction, namely $\mathcal{L}_1=\mathcal{L}^1_1\times \mathcal{L}^2_2$, where $\hat{\chi}_j(x,t)=\hat{\chi}_1(x-v_j t)$ and consequently
\begin{equation}
\mathcal{L}^j_1=\frac{1}{r} \langle e^{\frac{-i}{\sqrt{2}}\hat{\chi}_j(x_1-v_j t_1)} e^{\frac{-i}{\sqrt{2}}\hat{\chi}_j(x_2-v_j t_2)} e^{\frac{i}{\sqrt{2}}\hat{\chi}_j(x_3-v_j t_3)} e^{\frac{i}{\sqrt{2}}\hat{\chi}_j(x_4-v_j t_4)}\rangle_0.
\end{equation}
Further in Gaussian approximation~\cite{Edvin4}, in term of new bosonic fields, the above correlation function takes the from
\begin{equation}
\label{Four point correlation function}
\mathcal{L}^j_1=\frac{1}{r}\exp\left[-\frac{1}{4}\sum^4_{i=1} \lambda^2_i \langle \hat{\chi}^2_j(x_i-v_j t_i)\rangle_0-\frac{1}{2}\sum^4_{i<l} \lambda_i \lambda_l \langle \hat{\chi}_j(x_i-v_j t_i) \hat{\chi}_j(x_l-v_j t_l)\rangle_0 \right], \quad \lambda_i=\pm 1.
\end{equation}
Further calculations gives the final result for correlation function
\begin{equation}
\mathcal{L}^j_1=\frac{1}{r}\sqrt{\frac{L_{13}L_{14}L_{23}L_{24}}{L_{12}L_{34}}}, \quad L_{ij}=\frac{-irT}{2v_j}\frac{1}{\sinh[\pi T(t_i-t_j-(x_i-x_j)/v_j-i\gamma)]}, \quad i,j=1,2,3,4.
\end{equation}
All other four-point correlation functions can be calculated in the same manner. Correlation function at zero temperature can be obtained using the Eq.~(\ref{Zero temperature. Two-point correlation function. Section A}). It is worth mentioning that the higher order correlation functions in the perturbative expansion, which give the sub-leading corrections to current and noise, may depend on the regularization of the point-like tunnelling Hamiltonian~\cite{Michelle}.
\section{Time invariance of vertex correlation functions}
\label{Sec:C}
In this Appendix we show the time invariance of vertex correlation functions, namely that
\begin{equation}
\label{Time invariance. Section C}
\langle \hat{A}^{\dagger}(t)\hat{A}(0) \rangle_0 =\langle \hat{A}(t)\hat{A}^{\dagger}(0)\rangle_0=\langle \hat{A}^{\dagger}(0)\hat{A}(-t) \rangle_0.
\end{equation}
To do this, we represent these vertex correlation function though two-point correlation functions $G_j(x-x^{\prime},t-t^{\prime})$ defined in previous Appendix. For filling factor $\nu=1$ it is obvious because of Wick's theorem. For filling factor $\nu=2$ but without interaction
\begin{equation}
\label{Vertex correlation function. Section C}
\begin{split}
& \langle \hat{A}^{\dagger}(t_1)\hat{A}(t_2)\rangle_0=\langle \hat{A}(t_1)\hat{A}^{\dagger}(t_2)\rangle_0=[p^2+(1-p)^2] \sum_{j=1,2} \left[G_j(0,t_1-t_2)G_j(0,t_1-t_2)-G_j(\xi,t_1-t_2)G_j(-\xi,t_1-t_2) \right]+ \\
& + p(1-p)\left[2G_1(0,t_1-t_2)G_2(0,t_1-t_2)-G_1(\xi,t_1-t_2)G_2(-\xi,t_1-t_2)-G_2(\xi,t_1-t_2)G_1(-\xi,t_1-t_2)\right],
\end{split}
\end{equation}
thus is it straightforward to confirm the time invariance relations. At filling factor $\nu=2$ in presence of interaction using the Eq.~(\ref{Four point correlation function}) and commutation relation $[\hat{\chi}_{\alpha} (x_1-v_{\alpha} t_1),\hat{\chi}_{\beta}(x_2-v_{\beta} t_2)]=i\pi \delta_{\alpha \beta} \text{sgn}(x_1-x_2-v_{\alpha} t_1+v_{\beta} t_2)$
one can justify the Eq.~(\ref{Time invariance. Section C}).

\end{widetext}

\bibliography{refs}

\end{document}